\documentclass[twocolumn,epsf,graphics,psfig,superscriptaddress]{revtex4}

\usepackage{graphicx,graphics} 
\usepackage[dvips,bookmarks=false]{hyperref}
\usepackage{graphicx}
\usepackage{dcolumn}
\usepackage{eucal}
\usepackage[dvips]{epsfig}

\begin{document}
\title{Quantum size effects on spin-tunneling time in a magnetic resonant tunneling diode}
\author{Alireza Saffarzadeh} \email{a-saffar@tehran.pnu.ac.ir}
\affiliation{Department of Physics, Payame Noor University,
Nejatollahi Street, 159995-7613 Tehran, Iran}
\affiliation{Computational Physical Sciences Laboratory,
Department of Nano-Science, Institute for Research in Fundamental
Sciences (IPM), P.O. Box 19395-5531, Tehran, Iran}
\author{Reza Daqiq}
\affiliation{Department of Physics, Islamic Azad University,
Firuzkuh Branch, Firuzkuh, Iran }
\date{\today}

\begin{abstract}

We study theoretically the quantum size effects of a magnetic
resonant tunneling diode (RTD) with a (Zn,Mn)Se dilute magnetic
semiconductor layer on the spin-tunneling time and the spin
polarization of the electrons. The results show that the
spin-tunneling times may oscillate and a great difference between
the tunneling time of the electrons with opposite spin directions
can be obtained depending on the system parameters. We also study
the effect of structural asymmetry which is related to the
difference in the thickness of the nonmagnetic layers. It is found
that the structural asymmetry can greatly affect the traversal
time and the spin polarization of the electrons tunneling through
the magnetic RTD. The results indicate that, by choosing suitable
values for the thickness of the layers, one can design a high
speed and perfect spin-filter diode.

\end{abstract}
\maketitle

\section{Introduction}
The field of semiconductor spintronics has attracted a great deal
of attention during the past decade because of its potential
applications in new generations of transistors, lasers, and
integrated magnetic sensors. In addition, magnetic resonant
tunneling diodes (RTDs) can also help us to more deeply understand
the role of spin degree of freedom of the tunneling electron and
the quantum size effects on spin transport processes \cite{Wolf}.
By employing such a magnetic RTD, an effective injection of
spin-polarized electrons into nonmagnetic semiconductors (NMSs)
can be demonstrated \cite{Petukhov1,Vurgaftman}. In this regard,
the II-VI diluted magnetic semiconductors (DMSs)
\cite{Schmidt1,Furdyna} are known to be good candidates for
effective spin injection into a NMS because their spin
polarization is nearly 100\% and their conductivity is comparable
to that of typical NMS. A very promising II-VI DMS for spin
injection is (Zn,Mn)Se, which has been previously used for spin
injection experiments into GaAs \cite{Fiederling} and ZnSe
\cite{Schmidt2}. The (Zn,Mn)Se-based RTD with highly
spin-polarized electron current has been suggested by Egues
\cite{Egues} and experimentally demonstrated by Slobodskyy et al.
\cite{Slob1}. Also, different types of magnetic RTDs have been
proposed both theoretically
\cite{Beletskii,Havu,Li,Rad,Ertler2,Ertler1,Qiu} and
experimentally \cite{Oiwa,Maximov,Ohya,Fang,Slob2,Iovan}.

One of the key parameters in operation of magnetic RTDs is the
time aspect of tunneling process, which has been the focus of much
research in the past decade, because it is an important parameter
for better understanding of the spin-dependent tunneling phenomena
in high-speed devices. This quantity may strongly depend on the
quantum size of the devices, however, to the best of our
knowledge, no theoretical study on the dependence of quantum size
on the spin-tunneling time in the magnetic RTDs has so far been
reported.

Recently, based on the group velocity concept, several theoretical
studies of tunneling time in different magnetic junctions have
been done. Guo \textit{et al.} found obvious features of
separation of spin-tunneling time in ZnSe/Zn$_{1-x}$Mn$_x$Se
heterostructures \cite{Guo1,Guo2,Guo3}. Zhai \textit{et al.}
\cite{Zhai} studied the tunneling time in magnetic barrier
structures consisting of two identical or unidentical magnetic
barriers and magnetic wells. Wang \textit{et al.} \cite{Wang}
investigated tunneling properties of spin-polarized electrons
traversing ferromagnetic/insulator (semiconductor) double
junctions and reported that the tunneling time strongly depends on
the spin orientation of tunneling electrons. The effects of Rashba
spin-orbit interaction \cite{Wu,Zhang1} and Dresselhaus spin-orbit
coupling \cite{Zhang2} on the traversal time of ferromagnetic/
semiconductor/ferromagnetic heterostructures have also been
investigated. By considering the Rashba spin-orbit coupling in the
semiconductor and significant quantum size simultaneously, it has
been found that, as the length of the semiconductor increases, the
spin-tunneling time will increase with a behavior of slight
oscillation, whether for the spin-up electrons or for the
spin-down ones \cite{Zhang1}. Furthermore, it has been
demonstrated that the Dresselhaus spin-orbit coupling, unlike the
Rashba spin-orbit interaction that damps the motion of electrons,
does not prolong the traversal time of electrons tunneling through
the heterostructures \cite{Zhang2}.

The aim of the present study is to investigate the quantum size
effects of a typical magnetic RTD on spin-tunneling time and its
dependence on the concentration of magnetic ions. Our device,
similar to that used in Ref.\cite{Slob1}, is based on a quantum
well made of diluted magnetic semiconductor (Zn,Mn)Se between two
(Zn,Be)Se barriers and surrounded by highly n-type ZnSe layers. In
such a structure, the energy levels of the quantum well states
depend on spin direction due to the exchange splitting, meaning
that the energy levels of the (Zn,Mn)Se layer for spin-up
electrons will be different from that of spin-down electrons. We
will show here that this spin splitting of the energy levels,
which is controlled by an applied magnetic field and the value of
Mn concentration, enables one to select the resonant condition for
the desired spin by adjusting the quantum size of the device. This
paper is organized as follows. In Sec. II, we present the model
and formalism for traversal time of electrons through
ZnSe/ZnBeSe/ZnMnSe/ZnBeSe/ZnSe magnetic RTD. Numerical results and
discussions for spin-tunneling time and the degree of
electron-spin polarization in both the symmetric and the
asymmetric structures are presented in Sec. III. We conclude our
findings in Sec. IV.

\section{Model and formalism}
Consider a spin unpolarized electron current injected into a
ZnSe/ZnBeSe/Zn$_{1-x}$Mn$_x$Se/ZnBeSe/ZnSe structure shown in Fig.
1, in the presence of magnetic and electric fields along the
growth direction (taken as the $z$-axis). The conduction electrons
that contribute to the electric currents interact with the 3d$^5$
electrons of the Mn ions via the sp-d exchange interaction. Hence,
the external magnetic field, $B$, gives rise to the spin splitting
of the conduction band states in the Zn$_{1-x}$Mn$_x$Se layer.
Therefore, the injected electrons see a spin-dependent potential.
Due to the absence of any kind of scattering center for the
electrons, the motion along the $z$-axis is decoupled from that of
the $x-y$ plane, which is quantized in the Landau levels with
energies $E_n=(n+\frac{1}{2})\hbar\omega_c$, where $n=0, 1,
2,\cdots$ and $\hbar\omega_c=eB/m^*$. In such a case, the motion
of electrons along the $z$-axis can be reduced to the following
one-dimensional Schr\"{o}dinger equation
\begin{equation}\label{HH}
-\frac{\hbar^2}{2m^*}\frac{d^2\psi_{\sigma_z}(z)}{dz^2}+U_{\sigma_z}(z)\psi_{\sigma_z}(z)
=E_z\psi_{\sigma_z}(z) \ ,
\end{equation}
where the electron effective mass $m^*$ is assumed to be identical
in all the layers, $E_z$ is the longitudinal energy of electrons,
$U_{\sigma_z}(z)$ is the effective potential seen by a traverse
electron and is given as $U_{\sigma_z}(z)=V_s+U_0-eV_az/L$ in the
ZnBeSe layers ($0<z< L_1$ and $L_1+L_2<z< L$) where $U_0$ is the
height of the ZnBeSe barriers and
$U_{\sigma_z}(z)=V_s+V_x(z)+V_{\sigma_z}(z)-eV_az/L$ in the ZnMnSe
layer ($L_1<z< L_1+L_2$). Here, $L_1$ and $L_3$ are, respectively,
the widths of left and right ZnBeSe layers, and $L_2$ is the width
of the ZnMnSe layer ($L=L_1+L_2+L_3$);
$V_s=\frac{1}{2}g_s\mu_B{\bf\sigma}\cdot{\bf B}$ describes the
Zeeman splitting of the conduction electrons, where ${\bf\sigma}$
is the conventional Pauli spin operator; $V_x(z)$ is the
heterostructure potential or the conduction band offset in the
absence of a magnetic field, which depends on the Mn concentration
$x$ and is the difference between the conduction band edge of the
ZnMnSe layer and that of the ZnSe layer; $V_{\sigma_z}(z)$ is the
sp-d exchange interaction between the injected electron and the Mn
ions and can be calculated within the mean field approximation.
Hence, the sum of the last two terms can be written as
\begin{eqnarray}
&&V_x(z)+V_{\sigma_z}(z)\nonumber\\
&&=\left[\frac{1}{2}\Delta E(x)-N_0\alpha\sigma_zx_{\mathrm{eff}}SB_S\left(\frac{5\mu_BB}{k_B(T+T_0)}\right)\right]\nonumber\\
&&\times\Theta(z-L_1)\Theta(L_1+L_2-z)\ ,
\end{eqnarray}
where
\begin{eqnarray}\label{Ex}
\Delta E(x)&=&E_g(x)-E_g(0)\nonumber\\
&=&-0.63x+22x^2-195x^3+645x^4 \ ,
\end{eqnarray}
is the sum of the conduction and valance band offset under zero
magnetic field, when the real (effective) Mn concentration is $x$
($x_\mathrm{eff}=x[1-x]^{12}$). Here, $B_S(\cdots)$ is the
Brillouin function and $S=\frac{5}{2}$ is the spin of the Mn ions.
$\sigma_z=\pm \frac{1}{2}$ (or $\uparrow$, $\downarrow$) are the
electron-spin components along the magnetic field. We should note
that this form of Eq. \ref{Ex} is valid only for $0\leq x\leq 0.1$
\cite{Saffar1}. The last term in $U_{\sigma_z}(z)$ denotes the
effect of an applied bias $V_a$ along the $z$ axis on the system.

In order to study the tunneling time of electrons through the
structure, we adopt the group velocity approach
\cite{Anwar,Dragoman}, in which the tunneling time of a
spin-polarized electron can be defined as $\tau_{\sigma_z}=\int
dz/v_{g,\sigma_z}(z)$, where the spin-dependent group velocity,
$v_{g,\sigma_z}$, is defined as the ratio of the average
probability current density
$S_{\sigma_z}=\mathrm{Re}[\hbar(\psi_{\sigma_z}d\psi_{\sigma_z}/dz)/im^*]$
to the probability density $|\psi_{\sigma_z}|^2$ of the particle
\cite{Guo1,Guo2,Wang}. In this regard, the spin-tunneling time can
be written as
\begin{equation}\label{time}
\tau_{\sigma_z}=\frac{1}{\hbar}\int_0^L\frac{m^*}{|\gamma_{\sigma_z}(z)|\mathrm{Im}[\mathrm{tan}\theta_{\sigma_z}(z)]}dz\
,
\end{equation}
where
\begin{equation}\label{tan}
\mathrm{tan}\theta_{\sigma_z}(z)=\frac{1}{|\gamma_{\sigma_z}(z)|}
\left[\frac{1}{\psi_{\sigma_z}(z)}
\frac{d\psi_{\sigma_z}(z)}{dz}\right] \ ,
\end{equation}
\begin{equation}
\gamma_{\sigma_z}(z)=\frac{i}{\hbar}
\sqrt{2m^*[E_z-U_{\sigma_z}(z)]}\ .
\end{equation}
In the above equations, $\psi_{\sigma_z}(z)$ is the spin-dependent
wave function of the heterostructure. Since we have considered
that the electrons tunnel through the magnetic structure from the
left ($z<0$) to the right ($z>L$), under the influence of the
applied voltage $V_a$, the wave functions in each region can be
written as
\begin{eqnarray}\label{psi}
&&\psi_{\sigma_z}(z)=\nonumber\\
&&\left\{\begin{array}{cc}
e^{ik_{1\sigma_z}z}+r_{\sigma_z}e^{-ik_{1\sigma_z}z}, & z<0 ,\\
A_{2\sigma_z}{\rm Ai}[\rho_{\sigma_z}(z)]+B_{2\sigma_z}{\rm Bi}[\rho_{\sigma_z}(z)] , & 0<z<L_1 ,\\
A_{3\sigma_z}{\rm Ai}[\rho_{\sigma_z}(z)]+B_{3\sigma_z}{\rm Bi}[\rho_{\sigma_z}(z)] , & L_1<z<L_1+L_2 ,\\
A_{4\sigma_z}{\rm Ai}[\rho_{\sigma_z}(z)]+B_{4\sigma_z}{\rm Bi}[\rho_{\sigma_z}(z)] , & L_1+L_2<z<L ,\\
t_{\sigma_z}e^{ik_{5\sigma_z}z}, & z>L .\\
\end{array}\right.\nonumber\\
\end{eqnarray}
Here, $r_{\sigma_z}$ and $t_{\sigma_z}$ are the reflection and the
transmission amplitudes;
$k_{1\sigma_z}=\sqrt{2m^*(E_z-V_s)}/\hbar$ and
$k_{5\sigma_z}=\sqrt{2m^*(E_z-V_s+eV_a)}/\hbar$ are the electron
momenta; Ai[$\rho_{\sigma_z}(z)$] and Bi[$\rho_{\sigma_z}(z)$] are
the Airy functions with
$\rho_{\sigma_z}(z)=[E_z-U_{\sigma_z}(z)]L/(eV_a\lambda)$ and
$\lambda=[-\hbar^2L/(2m^*eV_a)]^{1/3}$.

The constants $A_{j\sigma_z}$ and $B_{j\sigma_z}$ (with $j$=2-4)
can be determined from a system of equations formed by
$\psi_{\sigma_z}$ and its derivative for the same $z$ value. Then,
we can relate $\psi_{\sigma_z}$ and its derivative
$\psi_{\sigma_z}'$ at two positions $z$ and $z'$ by the transfer
matrix ($M$) as follows:
\begin{eqnarray}\label{TM}
\left(\begin{array}{cc} \psi_{\sigma_z}'(z)\\\psi_{\sigma_z}(z)
\end{array}\right)
=\left(\begin{array}{cc}
M_{11}&M_{12}\\
M_{21}&M_{22}\end{array}\right) \left(\begin{array}{cc}
\psi_{\sigma_z}'(z')\\\psi_{\sigma_z}(z')
\end{array}\right) \ .
\end{eqnarray}

Applying the results of Eq. \ref{TM} in Eq. \ref{tan}, the value
of $\mathrm{tan}\theta_{\sigma_z}(z)$, which determines the
tunneling time at position $z$, can be written in terms of its
value at position $z'$ as
\begin{equation}\label{tanzz}
\mathrm{tan}\theta_{\sigma_z}(z)=\frac{1}{|\gamma_{\sigma_z}(z)|}
\frac{M_{11}|\gamma_{\sigma_z}(z')|\mathrm{tan}\theta_{\sigma_z}(z')+M_{12}}
{M_{21}|\gamma_{\sigma_z}(z')|\mathrm{tan}\theta_{\sigma_z}(z')+M_{22}}\
.
\end{equation}

Therefore, using Eqs. \ref{tanzz} and \ref{time}, the
spin-tunneling time can be calculated for the desired magnetic
RTD.
\begin{figure}
\centerline{\includegraphics[width=0.8\linewidth]{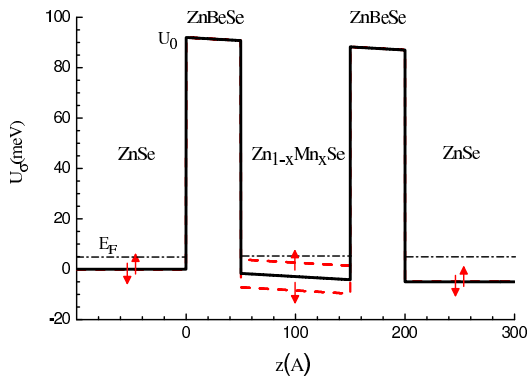}}
\caption{(Color online) Spin-dependent effective potential profile
in the magnetic RTD under influence of both the magnetic and
electric fields along the $z$ axis. The up and down arrows
represent spin-up and spin-down subbands, respectively. The two
ZnBeSe layers have the same widths $L_1=L_3$=50 {\AA}. $L_2$=100
{\AA}, $x$=0.04, $B$=2 T and $V_a$=5 mV.}
\end{figure}

\begin{figure}
\centerline{\includegraphics[width=0.8\linewidth]{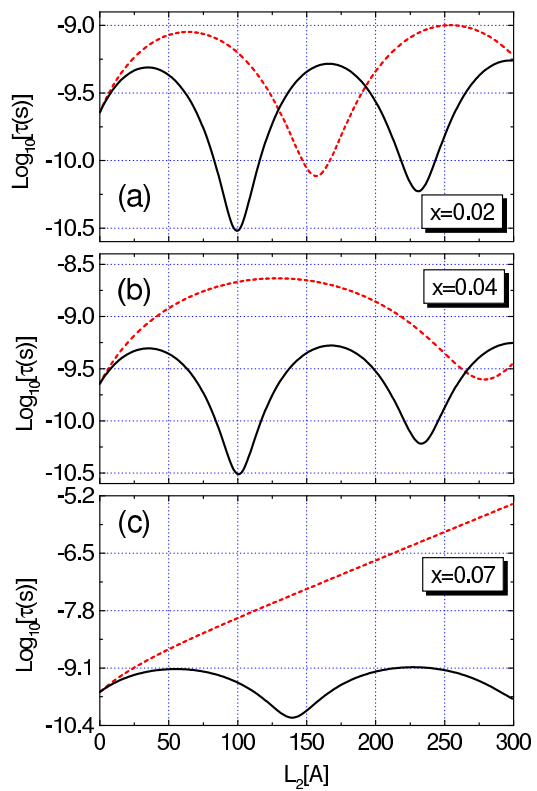}}
\caption{(Color online) Spin-tunneling time for electrons
traversing the symmetric structure: $L_1=L_3=50$ {\AA}. The dashed
and solid lines correspond to spin-up and spin-down electrons,
respectively.}
\end{figure}

\begin{figure}
\centerline{\includegraphics[width=0.8\linewidth]{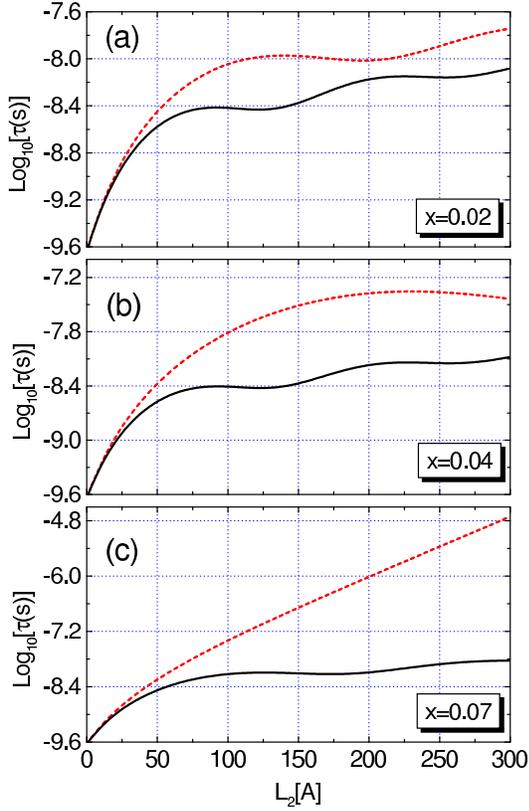}}
\caption{(Color online) Spin-tunneling time for electrons
traversing the asymmetric structure: $L_1=10$ {\AA} and $L_3=90$
{\AA}. The dashed and solid lines correspond to spin-up and
spin-down electrons, respectively.}
\end{figure}
\section{Results and discussion}
In this section, we use the formulas given above to investigate
the quantum size effect on the spin-tunneling time and
electron-spin polarization in the
ZnSe/ZnBeSe/Zn$_{1-x}$Mn$_x$Se/ZnBeSe/ZnSe heterostructures. In
the numerical calculations we have taken the following values:
$U_0=92$ meV \cite{Havu}, $T=2.2$ K, $T_0=1.4$ K, and
$N_0\alpha=-0.27$ eV \cite{Dai}, $E_F=5$ meV, $B=2$ T, $V_a=5$ mV,
$g_s=1.1$, and $m^*=0.16$ $m_e$ ($m_e$ is the mass of the free
electron). In the tunneling process at low temperatures, the
electrons with energy near Fermi energy ($E_F$) carry most of the
current; for this reason we have done our numerical calculations
at $E_z=E_F$. Also, due to the band-gap bowing of the
Zn$_{1-x}$Mn$_x$Se layer \cite{Dai,Saffar1}, we have examined the
effects of three values of Mn concentrations ($x=0.02$, 0.04, and
0.07) for the paramagnetic layer.

First, we study the tunneling time in the symmetric and the
asymmetric structures, depending on the thickness of the ZnBeSe
layers. The structure is called symmetric (asymmetric) if
$L_1=L_3$ $(L_1\neq L_3)$. Figure 2 shows the spin-tunneling time
as a function of thickness $L_2$ of the Zn$_{1-x}$Mn$_x$Se layer.
When $L_2$ is zero, the tunneling time is independent of the spin
orientation and hence $\tau_\uparrow=\tau_\downarrow$. At $x=0.02$
[Fig. 2(a)] and with increasing $L_2$, the tunneling time
oscillates for both spin-up and spin-down electrons; however, the
length period of oscillations for $\tau_\downarrow$ is shorter
than $\tau_\uparrow$. At this concentration, the paramagnetic
layer has its minimum value of band-gap, which is smaller than the
band gap of the ZnSe layers \cite{Saffar1}. In the case of
$x=0.04$ [Fig. 2(b)], the oscillation does not change considerably
for spin-down electrons, but the length period significantly
increases for the spin-up ones. At $x=0.07$, the tunneling time
oscillates only for spin-down electrons, as shown in Fig. 2(c). It
is important to note that at $x\simeq 0.04$ and under zero
magnetic field, the band gap of Zn$_{1-x}$Mn$_x$Se is nearly the
same as that of ZnSe, while for larger values of $x$ such as
$x=0.07$, the Zn$_{1-x}$Mn$_x$Se layer behaves as a potential
barrier in comparison with the ZnSe layers \cite{Dai}.
\begin{figure}
\centerline{\includegraphics[width=0.8\linewidth]{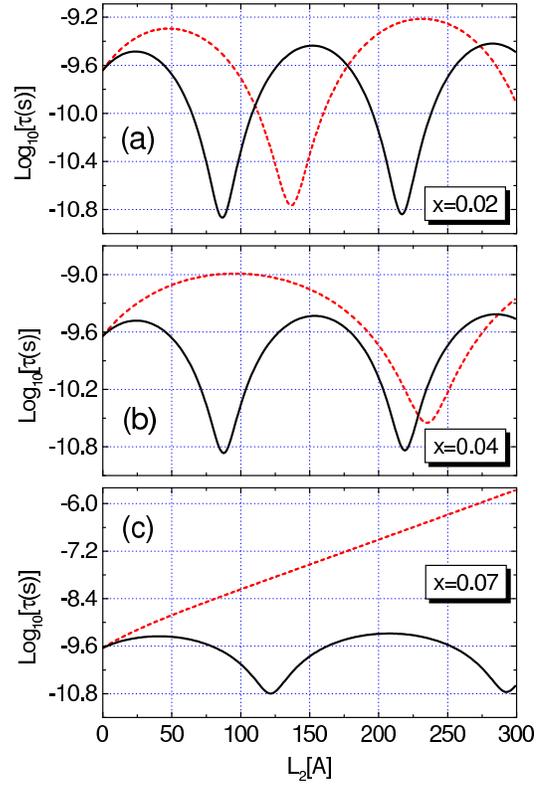}}
\caption{(Color online) Spin-tunneling time for electrons
traversing the asymmetric structure: $L_1=90$ {\AA} and $L_3=10$
{\AA}. The dashed and solid lines correspond to spin-up and
spin-down electrons, respectively.}
\end{figure}
\begin{figure}
\centerline{\includegraphics[width=0.8\linewidth]{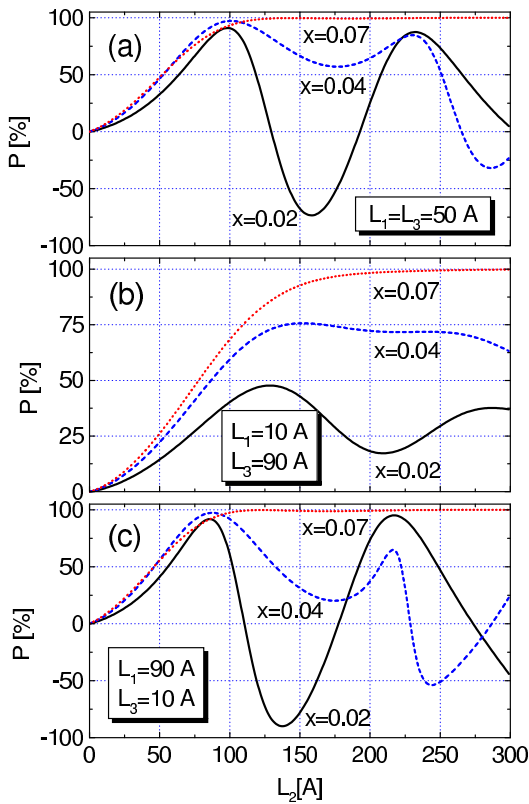}}
\caption{(Color online) Degree of spin polarization as a function
of the Zn$_{1-x}$Mn$_x$Se layer thickness $L_2$ for (a) symmetric
and [(b) and (c)] asymmetric structures.}
\end{figure}

The appearance of oscillatory and non-oscillatory behaviors in the
tunneling time is due to the effective potential $U_{\sigma_z}(z)$
and reflects this fact that the difference between the position of
the Fermi energy and the bottom of the conduction band on the
energy axis for each spin orientation in the paramagnetic layer
has dominant effect on the behavior of spin-tunneling time. The
physical origin of the oscillatory behavior is explained by the
quantum well states. As is well known for resonant tunneling
through double-barrier structures \cite{Guo4,Saffar2}, when the
incident energy of electrons coincides with the energy of a
quasibound state in the quantum well, a resonance condition is
fulfilled and the transmission coefficient of the electrons
through the heterostructure strongly increases. On the other hand,
the position of the quantum well states, formed in the
Zn$_{1-x}$Mn$_x$Se layer, strongly depends on the well thickness
$L_2$. Therefore, with continuous variation in $L_2$, the position
of the resonant states varies and this leads to the oscillations
of the tunneling time. As a remarkable feature in the
oscillations, one can see sharp dips in comparison with the broad
peaks in the tunneling time curves, which correspond to the
narrowing of the width of resonant bands arising from confinement
of electrons to the paramagnetic layer. With increasing $L_2$,
these narrow bands (levels) quickly cross the energy of incident
electrons in the left ZnSe layer on the energy axis, the tunneling
time decreases, and the sharp dips appear. In contrast, the peaks
are broad, due to the broad gaps between the discrete resonant
levels in the ZnMnSe layer.

Figs. 3 and 4 show the spin-tunneling times in two asymmetric
structures: (i) case $L_1<L_3$ where $L_1=10$ {\AA} and $L_3=90$
{\AA}; (ii) case $L_1>L_3$ where $L_1=90$ {\AA} and $L_3=10$
{\AA}. In the case of $L_1<L_3$ (Fig. 3), the spin-tunneling times
are almost increasing functions of $L_2$ and there is no
oscillation between $\tau_\uparrow$ and $\tau_\downarrow$ in all
three Mn concentrations. In contrast, for the case of $L_1>L_3$
(Fig. 4), the spin-tunneling times show oscillatory behavior very
similar to those of the symmetric structure ($L_1=L_3$). The
reason for discrepancy in the tunneling time of two asymmetric
structures is that the voltage drop within the ZnBeSe and ZnMnSe
layers depends on the position $z$ [see the last term in
$U_{\sigma_z}(z)$], and hence, the depth of quantum well in the
paramagnetic layer strongly depends on the position of the layer
with respect to the origin. Therefore, in the case of $L_1>L_3$,
the incident electrons see a deeper quantum well with respect to
the the case of $L_1<L_3$, and this affects the features of the
resonant states. If we increase the bias voltage, the oscillatory
behavior in the asymmetric structures with $L_1<L_3$ appears, too.
These results may be important from experimental point of view and
indicate that special care must be taken during sample growth in
order to make a magnetic RTD with low power consumption, high
speed, and greater spin-filter efficiency \cite{Ertler2}. It is
necessary to point out that the effects of the width of the
paramagnetic layer on the spin-dependent current densities in such
a magnetic RTD have been studied and the oscillatory behaviors in
the current-voltage characteristics reported \cite{Havu}.

Now, for further understanding of the quantum size effects of the
system on the spin-tunneling times, we calculate the degree of
spin polarization of the tunneling electrons, which can be defined
as $P=(\tau_\uparrow- \tau_\downarrow)/(\tau_\uparrow+
\tau_\downarrow)$. The results are plotted in Fig. 5 for both the
symmetric and the asymmetric structures. It is clear that the
value of spin polarization can be greatly changed by the Mn
concentration, the thickness of the ZnMnSe layer as well as the
status of structural symmetry. At $x=0.07$ and with increasing
$L_2$, the spin polarization retains positive and its value
increases almost linearly and reaches nearly 100\%, indicating an
excellent spin filtering effect. This means that for this value of
$x$ and in all the structures, the tunneling process of spin-up
(spin-down) electrons is always a slow (quick) process. In the
cases of $x=0.02$ and $x=0.04$, however, the spin polarization can
change sign for particular ranges of $L_2$ in the structures with
$L_1=L_3$ and $L_1>L_3$. Therefore, the case of
$\tau_\uparrow>\tau_\downarrow$ or $\tau_\uparrow<\tau_\downarrow$
may strongly depend on the thickness of the paramagnetic layer.

According to the above results, the tunneling process of the
spin-polarized electrons through the magnetic RTD can be divided
into slow and quick processes. However, we cannot say which one of
the spin orientations of the tunneling electrons always
corresponds to the slow process and which one corresponds to the
quick process. Such a feature occurs for Mn concentration $x\leq$
0.04 in both the symmetric and the asymmetric ($L_1>L_3$)
structures. We would like to point out here that our calculations
have been performed under the assumption of a phase-coherent
tunneling process, which applies to heterostructures with narrow
wells and barriers. When the heterostructures become thicker, we
should replace the phase-coherent tunneling by a sequential
process and these features will change.

\section{Conclusion}
Using the group velocity concept and the particle current
conservation principle, we have shown how the geometry and the
size of the device affect the spin-tunneling time and the degree
of spin polarization of tunneling electrons in a (Zn,Mn)Se-based
magnetic RTD. The tunneling time for spin-up and spin-down
electrons and hence the degree of spin polarization may strongly
depend on the width of the paramagnetic layer and the Mn
concentration. We found that, due to the oscillatory behavior of
spin-tunneling time with increasing thickness of the (Zn,Mn)Se
layer, special care should be taken for designing a magnetic RTD
with high efficiency. Furthermore, the present results may open a
new way to control the degree of spin polarization and design the
high speed magnetic devices.

\end{document}